\documentclass[pre,twocolumn,showpacs,amsmath,amssymb]{revtex4}
\usepackage{bm}       
\usepackage[dvips]{graphicx}
\usepackage{color}

\newcommand{\beq}{\begin{equation}}
\newcommand{\eeq}{\end{equation}}
\newcommand{\Matlab}{\textsc{Matlab}}

\newcommand{\g}{\gamma}
\newcommand{\gO}{\gamma_{\Omega}}
\newcommand{\bfu}{\mathbf{u}}
\newcommand{\bfug}{\bfu(\gamma)}
\newcommand{\bfr}{\mathbf{r}}

\newcommand{\bfnabla}{\boldsymbol{\nabla}}

\newcommand{\ain}{a_0}
\newcommand{\aout}{a_L}
\newcommand{\ug}{u(\gamma)}
\newcommand{\kg}{k(\gamma)}
\newcommand{\ugO}{u(\gO)}
\newcommand{\kgO}{k(\gO)}
\newcommand{\ka}{k_a}
\newcommand{\Kx}{K(x)}
\newcommand{\KL}{K(L)}
\newcommand{\mk}{k_{\Omega}}

\newcommand{\conv}{\mathcal{C}}
\newcommand{\Damk}{\delta}
\newcommand{\Damknr}{Damk\"{o}hler number}

\newcommand{\Damkb}{\Damk_{\beta}}

\definecolor{newc}{rgb}{0,0,0.3}

\begin{document}

\title{Scaling behavior of optimally structured catalytic microfluidic reactors}

\author{Fridolin Okkels and Henrik Bruus}

\affiliation{MIC -- Department of Micro and Nanotechnology,
NanoDTU, Technical University of Denmark, Bldg 345 East, DK-2800
Lyngby, Denmark}

\date{30 June 2006}

\begin{abstract}
In this study of catalytic microfluidic reactors we show that,
when optimally structured, these reactors share underlying scaling
properties. The scaling is predicted theoretically and verified
numerically. Furthermore, we show how to increase the reaction
rate significantly by distributing the active porous material
within the reactor using a high-level implementation of topology
optimization.
\end{abstract}


\maketitle

Chemical processes play a key role for the production and analysis
of many substances and materials needed in industry and heath
care. Generally, the optimization of these processes is an
important goal, and with the introduction of microfluidic reactors
involving laminar flows, the resulting concentration distributions
mean better control and utilization of the reactors
\cite{KiwiMinsker05a}. These conditions make it possible to design
reactors using the method of topology optimization
\cite{SigmundBook}, which recently has been applied to fluidic
design of increasing complexity
\cite{Borrvall:03a,AGH-fluid,Olesen:06a}.

First, we report the finding of scaling properties of such optimal
reactors. To illustrate the method we study a simple model of a
chemical reactor, in which the desired product arises from a
single first-order catalytic reaction due to a catalyst
immobilized on a porous medium filling large regions of the
reactor.

Next, we show that topology optimization can be employed to design
optimal chemical micro-reactors. The goal of the optimization is
to maximize the mean reaction rate of the micro-reactor by finding
the optimal porosity distribution of the porous catalytic support.
Despite the simplicity of the model, our work shows that topology
optimization of the design of the porous support inside the
reactor can increase the reaction rate significantly.

Our model system is a first-order catalytic reaction,
$\mathrm{A}\stackrel{\mathrm{C}}{\longrightarrow} \mathrm{B}$,
taking place inside a microfluidic reactor of length $L$,
containing a porous medium of spatially varying porosity
$\g(\bfr)$ and a buffer fluid filling the pores. The porosity $\g$
is defined as the local volume fraction occupied by the buffer
fluid \cite{Desmet:03a}, and it can vary continuously from zero to
unity, where $\g = 0$ is the limit of dense material (vanishingly
small pores) and $\g = 1$ is the limit of pure fluid (no porous
material). The reactant A and the product B are dissolved with
concentrations $a$ and $b$, respectively, in the buffer fluid,
which is driven through the reactor by a constant, externally
applied pressure difference $\Delta p$ between an inlet and outlet
channel. The catalyst C is immobilized with concentration $c$ on
the porous support.

The working principle of the reactor is quite simple. The buffer
fluid carries the reactant A through the porous medium supporting
the catalyst C. The reaction rate is high if at the same time the
reactant A is supplied at a high rate and the amount of
immobilized catalyst C is large. However, these two conditions are
contradictory. For a given pressure drop $\Delta p$ the supply
rate of A is high if $\g$ is high allowing for a large flow rate
of the buffer fluid. Conversely, the amount of catalyst C is high
if $\g$ is low corresponding to a dense porous support with a
large active region. Consequently, an optimal design of the porous
support must exist involving intermediate values of the porosity.
Besides, the optimal design may involve an intricate distribution
of porous support within the reactor, and to find this we employ
the method of topology optimization in the implementation of
Ref.~\cite{Olesen:06a}.

In the steady-state limit, the reaction kinetics is given by the
following advection-diffusion-reaction equation for the reactant
concentration $a$,
 \beq \label{eq:ChemAdvDiff}
 \big[\bfug\cdot\bfnabla\big] a = D \bfnabla^2 a - \kg\,a.
 \eeq
Here $\bfug$ is the velocity field of the buffer fluid, $D$ is the
diffusion constant of the reactant in the buffer, and $-\kg\,a$ is
the reaction term of the first order isothermal reaction, which
depends on the concentration of the catalyst C through $\g(\bfr)$.
In this problem three characteristic timescales $\tau^{{}}_A,\
\tau^{{}}_R$ and $\tau^{{}}_D$ naturally arise,
 \beq \label{eq:CharTimeSc}
 \tau^{{}}_A = \frac{L}{u},\quad \tau^{{}}_D = \frac{L^2}{D},\quad \tau^{{}}_R = \frac{1}{\mk},
 \eeq
which correspond directly to the advection, diffusion, and
reaction term in Eq.~(\ref{eq:ChemAdvDiff}), respectively. These
time-scales will be used in the following theoretical analysis.
Note that the index of $\Omega$ generally denote an average over
the design region, e.g., $\mk=\langle \kg \rangle_{\Omega}$.

The porosity field $\g(\bfr)$ uniquely characterizes the reactor
design since it determines both the distribution of the catalyst
and the flow of the buffer. In the Navier--Stokes equation,
governing the flow of the buffer, the presence of the porous
support can be modelled by a Darcy damping force density
$-\alpha(\g)\,\bfu$, where $\alpha$ is the local,
porosity-dependent, inverse permeability\cite{Bear72}. Assuming
further that the buffer fluid is an incompressible liquid of
density $\rho$ and dynamic viscosity $\eta$, the governing
equations of the buffer in steady-state become
 \begin{subequations}
 \begin{align}
 \label{eq:NS}
 \rho (\bfu\cdot\bfnabla)\bfu & = -\bfnabla p + \eta
 \nabla^2\bfu
 -\alpha(\g)\,\bfu, \\
 \label{eq:ContEq}
 \bfnabla\cdot\bfu & = 0.
 \end{align}
 \end{subequations}
The coupling between $\alpha$ and $\g$ is given by the function
\mbox{$\alpha(\gamma) \equiv \alpha_{\max}\,
\frac{q(1-\gamma)}{q+\gamma}$}, where $\alpha_{\max}$ is
determined by the non-dimensional Darcy number $Da =
\frac{\eta}{\alpha_{\max}\,L^2}$, and $q$ is a positive parameter
used to ensure global convergence of the topology optimization
method\cite{Borrvall:03a,Olesen:06a}. In this work $Da$ is
typically around $10^{-5}$, resulting in a strong damping of the
buffer flow inside the porous support. The model is solved for a
given $\g(\bfr)$ by first finding $\bfug$ from Eqs.~(\ref{eq:NS})
and (\ref{eq:ContEq}) and then $a(\bfr)$ from
Eq.~(\ref{eq:ChemAdvDiff}).

Our aim is to optimize the average reaction-rate
$(\kg\,a)_{\Omega}$ of the reactor by finding the optimal porosity
field $\g(\bfr)$. We therefore introduce the following objective
function $\Phi(\g)$, which by convention has to be minimized,
 \beq \label{eq:ObjFunc}
 \Phi(\g) = -(\kg\,a)_{\Omega}.
 \eeq

To better characterize the performance of the reactor and to
introduce the related quantities, we first analyze a simple 1D
model defined on the $x$-axis. The porous medium is placed in the
reaction region $\Omega$ extending from $x=0$ to $x=L$.
Eq.~(\ref{eq:ContEq}) leads to a constant flow velocity $u$, and
as the complete pressure-drop occurs in the porous medium, we have
$p(0)=p_0+\Delta p$ and $p(L)=p_0$. In this case the boundary
conditions for the advection-diffusion-reaction equation
Eq.~(\ref{eq:ChemAdvDiff}) are $a(-\infty)=\ain$, $a'(-\infty) =
0$, and $a'(\infty) = 0$, where the primes indicate
$x$-derivatives. We denote the outlet concentration
$a(\infty)=\aout$. From Eqs.~(\ref{eq:ChemAdvDiff})
and~(\ref{eq:ObjFunc}), we then derive the following expression of
the objective function

 \begin{align}
 \Phi(\g) & = -\langle \kg\,a \rangle
 = \frac{1}{L}\int_0^L\!\!
 \left[ \ug a' - D\,a'' \right] dx \nonumber \\
 & = \frac{\ug}{L}\left( \aout - \ain \right) -
 \frac{D}{L}\big[a'(L) - a'(0)\big].
 \label{eq:1DPhi1}
 \end{align}

For simplicity, we now limit the analysis to the non-diffusive
case ($D=0$), and from Eq.~(\ref{eq:1DPhi1}) we get the objective
function defined in terns of the reaction conversion $\conv$,
 \beq \label{eq:1DMeanRate2}
 \Phi(\g)  = -\frac{\ug\,\ain}{L}\,\conv, \quad \mathrm{with}
 \quad \conv \equiv 1-\frac{\aout}{\ain}.
 \eeq

With an explicit $x$ dependence of the reaction rate coefficient
$k(x)$, we obtain $\ug a' = - k(x)\,a$ with the solution
 \beq \label{eq:1DcSol}
 a(x) = \ain\,e^{-\frac{\Kx}{\ug}},\quad \mathrm{with} \quad
 \Kx \equiv \int_0^x k(\tilde{x}) d\tilde{x}.
 \eeq
This leads to the following expression of the conversion:
 \beq \label{eq:1DcSolOut}
 \conv = 1-e^{-\frac{\KL}{\ug}} = 1-e^{-\frac{\mk}{\ug}L}
       = 1-e^{-\Damk},
 \eeq
where we have introduced the dimensional-less \Damknr{} $\Damk$
\cite{Damkohler:1937}
 \beq \label{eq:1Dxidef}
 \Damk \equiv \frac{\tau^{{}}_A}{\tau^{{}}_R} = 
 \frac{\mk}{\ug}L,
 \eeq
having the physical interpretation of the ratio between the
advection and the reaction timescale.

To derive the flow speed $\ug$ in the 1D model we first let $q
\rightarrow \infty$ resulting in $\alpha(\g)=\alpha_{\max} (1-\g)$
and then by integrating Eq.~(\ref{eq:NS}), $0=-p' - \alpha_{\max}
(1-\g)\,u$, we get
 \beq 
 u = \frac{\Delta p}{\alpha_{\max}(1-\gO)L}
 = \frac{Da}{1-\gO}\,\frac{\Delta p\,L}{\eta}.
 \label{eq:1DVel}
 \eeq

To solve the 1D optimization problem analytically, we chose to
abandon the spatial variations of $\g$ in the 1D model. We have to
find the solutions to $\frac{\partial \Phi}{\partial \gO} \equiv
0$, and from Eqs.~(\ref{eq:1DMeanRate2}) and (\ref{eq:1DcSolOut}),
we end up by having to solve the following equation
 \beq \label{eq:1DOptRelation}
 1-e^{\Damk}+\Damk(1+\beta) = 0,
 \eeq
where we have assumed that $\kgO \propto (1-\gO)^{\beta}$. The
specific properties of the catalytic reaction determines the value
of $\beta$, e.g., if the full volume of the porous medium is
active then $\beta = 1$, while if only the surface is active then
$\beta = 2/3$. Solving Eq.~(\ref{eq:1DOptRelation}) gives the
optimal value of $\Damk_{\beta}$, where the reference to $\beta$
now is explicit.

All numerical solutions are found using the commercial numerical
modelling-tools \Matlab \cite{MATLAB} and COMSOL \cite{COMSOL}. To
validate numerically the analytic results of the 1D model, we
solve Eqs.~(\ref{eq:ChemAdvDiff}) and (\ref{eq:1DVel}) for a given
homogeneous design variable $\gO$ and find the optimal value using
a brute-force optimization method \cite{MATLABOptim}.

To obtain a general scaling parameter for the problem defined in
Eq.~(\ref{eq:ChemAdvDiff}) we reintroduce diffusion. However, to
minimize the trivial influence from the inlet and outlet, we only
study the limit of low diffusion, e.g., $\tau^{{}}_D \gg
\tau^{{}}_A,\tau^{{}}_R$. In this limit the optimal reactors
involve a balance between the advection and reaction processes,
and consequently we expect that $\tau^{{}}_A$ and $\tau^{{}}_R$,
should enter on equal footing in the scaling parameter. We are
therefore led to propose the following dimensional-less form of
the scaling-parameter
 \beq \label{eq:1DdefPsi}
 \frac{\sqrt{\tau^{{}}_A\,\tau^{{}}_R}}{\tau^{{}}_D}
   = \frac{D}{\sqrt{\kgO\,\ugO\,L^3}}.
 \eeq

\begin{figure}[ht!] 
\begin{center}
\includegraphics[width=\columnwidth]{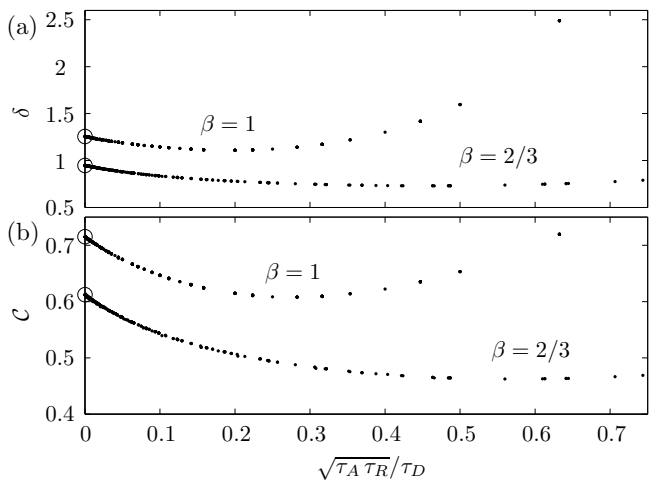}
\end{center}
\caption{Plot of 1D results showing (a) \Damknr{} $\Damk$ and (b)
conversion $\conv$, both as a function of
$\sqrt{\tau^{{}}_A\,\tau^{{}}_R}/\tau^{{}}_D$ and for optimal
choices of porosity. In both cases $\beta = 2/3,\,1$ and the
parameter-scan of each choice consist of 512
optimizations\cite{ParamScan} which collapse nicely. For zero
difffusion, $\sqrt{\tau^{{}}_A\,\tau^{{}}_R}/\tau^{{}}_D=0$,
Eqs.~(\ref{eq:1DOptRelation}) and (\ref{eq:1DcSolOut}) give the
exact results $\Damk_{2/3}=0.9474$, $\Damk_{1}=1.2564$,
$\conv_{2/3}=0.6122$,
 and $\conv_{1}=0.7153$, which are marked by circles on the ordinate
axis.} \label{fig:xiPlot1}
\end{figure}
%
\begin{figure}[ht!]
\begin{center}
\includegraphics[width=\columnwidth]{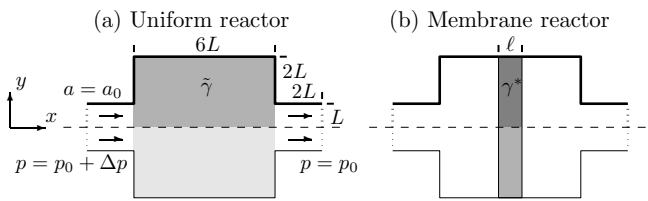}
\end{center}
\caption{Illustration of the two simple 2D reactor setups, (a) the
uniform reactor with porosity $\tilde{\g}$ and (b) the membrane
reactor of width $\ell$ and with porosity $\g^{\ast}\equiv 0$. The
horizontal dashed line is a symmetry-line.}\label{fig:Setup}
\end{figure}

Figure \ref{fig:xiPlot1} shows that the measured values of
$\Damkb$ and $\conv_{\beta}$ for optimal porosity both scale with
respect to $\sqrt{\tau^{{}}_A\,\tau^{{}}_R}/\tau^{{}}_D$. The
simulations cover 512 optimal reactors in a wide and dense
parameter-scan \cite{ParamScan}, and as they collapse almost
perfectly on single curves, we have not distinguished the
data-points further. In the non-diffusive case $D=0$ and
$\sqrt{\tau^{{}}_A\,\tau^{{}}_R}/\tau^{{}}_D = 0$, exact values of
$\Damkb$ and $\conv_{\beta}$ are determined by
Eqs.~(\ref{eq:1DcSolOut}) and (\ref{eq:1DOptRelation}), and they
match exactly with the numerical results, as seen in
Fig.~\ref{fig:xiPlot1}, where they are marked by circles on the
ordinate.

\begin{figure}[ht!]
\begin{center}
\includegraphics[width=\columnwidth]{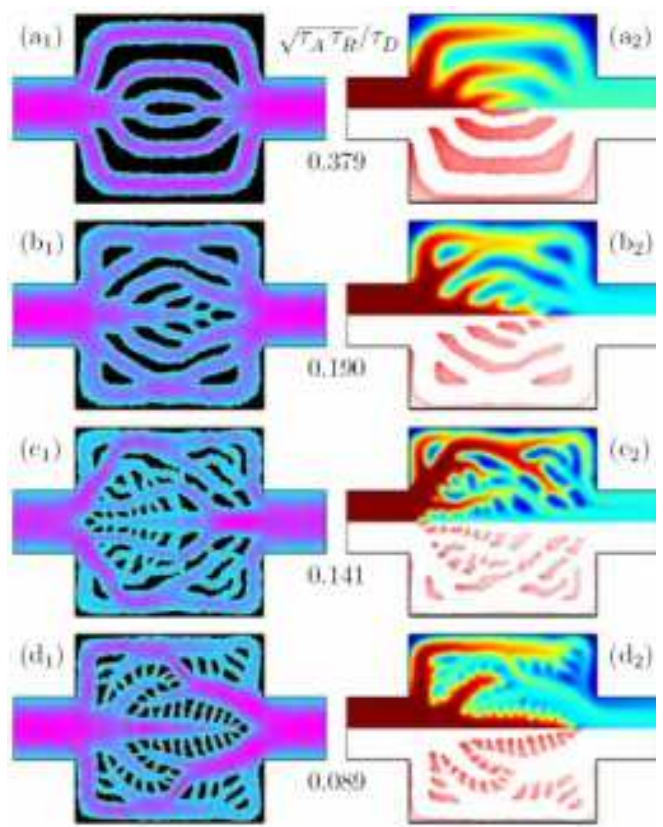}
\end{center}
\caption{Representative collection of topology optimized reactor
designs for deceasing values of
$\sqrt{\tau^{{}}_A\,\tau^{{}}_R}/\tau^{{}}_D$. (Left column) The
distribution $\gamma$ of porous material in black together with a
color-grading indication of the flow speed $u$. (Right column) The
concentration $a$ on top with the reaction rate $\kg a$ below.
Parameters (in SI units) $L=10^{-3}$ and the following values of
$[Da,\,D,\,\Delta p,\,k_a]$: (a)~$[10^{-4},\,3 \times
10^{-8},\,0.25,\,0.25]$, (b)~$[10^{-4},\,3 \times
10^{-8},\,0.25,\,1]$, (c)~$[10^{-5},\,10^{-8},\,0.5,\,1]$, and
(d)~$[10^{-4},\,10^{-8},\,0.25,\,0.5]$. \label{fig:TOPOPT} }
\end{figure}

We now introduce three types of 2D reactors: the uniform reactors,
Fig.~\ref{fig:Setup}(a), the membrane reactors,
Fig.~\ref{fig:Setup}(b), and the topology optimized reactors, for
which a few is shown in Fig.~\ref{fig:TOPOPT}. First we optimize
the simple reactors in Fig.~\ref{fig:Setup}. They both depend only
on one variable, which for the uniform reactor is the uniform
porosity $\tilde{\g}$, and for the membrane reactor is the width
$\ell$ of a porous membrane of porosity $\g^{\ast} \equiv 0$
\cite{MATLABOptim}. Because of mirror-symmetry in the $xz$-plane,
only the upper half of the reactors are solved in all the
following work.

In the third type of 2D reactors we let the porosity $\g(\bfr)$
vary freely within the same design region as for the uniform
reactor. The optimal design is found using the topology
optimization method, described in detail in
Ref.~\cite{Olesen:06a}. This is an iterative method, in which,
starting from an initial guess $\g_0$ of the design variable, the
$n$th iteration consists of first solving the systems for the
given design variable $\g_n$, then evaluating the sensitivities
$\frac{\partial \Phi}{\partial \g}$ by solving a corresponding
adjoint problem, and finally obtaining an improved updated
$\g_{n+1}$ by use of the "method of moving asymptotes"
(MMA)\cite{MMA,MMAprogram}. In Fig.~\ref{fig:TOPOPT} is shown a
representative collection of topology optimized designs together
with the corresponding flow speed $u$, concentration $a$, reaction
rate $\kg a$, and parameter values. In the large parameter space
under investigation, our work shows a systematic decrease of
pore-sizes and the emergence of finer structures in the topology
optimized reactors as the scaling parameter
$\sqrt{\tau^{{}}_A\,\tau^{{}}_R}/\tau^{{}}_D$ is decreased.

\begin{figure}[ht!]
\includegraphics[width=0.8\columnwidth]{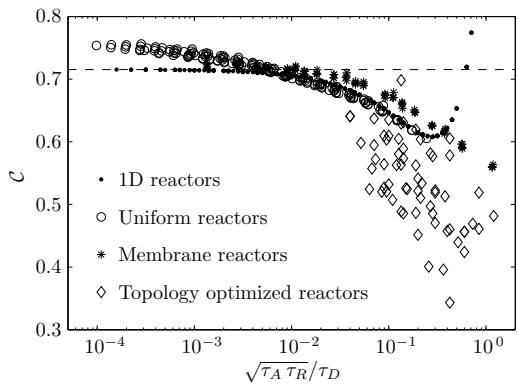}
\caption{Overall scaling of the conversion $\conv$ as a function
of $\sqrt{\tau^{{}}_A\,\tau^{{}}_R}/\tau^{{}}_D$ for the different
optimal reactors. The abscissa is logarithmic to emphasize the
common scaling behavior. The dashed line indicate the theoretical
value $\mathcal{C}_1$ for zero diffusion in the 1D case.
\label{fig:MultiPlotconv}}
\end{figure}

In Fig.~\ref{fig:MultiPlotconv} the conversion $\conv$ is plotted
as a function of $\sqrt{\tau^{{}}_A\,\tau^{{}}_R}/\tau^{{}}_D$ for
all optimal reactors of this work. It shows that all reactors
collapse on curves similar to the 1D reactors, although the
topology optimized reactors exhibit a larger spread. We believe
that this scaling is a signature of a general property of optimal
immobilized catalytic reactors. Note that the conversion of the
uniform reactor in the low diffusion limit is a few percent higher
than the theoretical estimate, an effect caused by low convection
in the corners, resulting in 'dead zones'.

%

\begin{figure}[ht!]
\includegraphics[width=0.9\columnwidth]{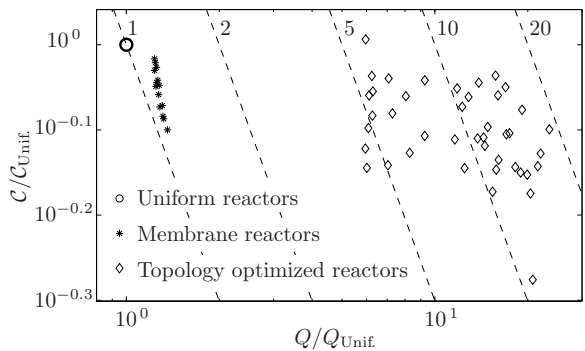}
\caption{Log-log plot of the relation between convection $\conv$
and flow rate $Q$ for the different reactors, when normalized with
values $\conv_{\mathrm{unif}}$ and $Q_{\mathrm{unif}}$ of the
uniform reactors. The reaction rate improvements are shown at the
top (see text).\label{fig:PhiCompare} }
\end{figure}

In terms of the objective function $\Phi$ the topologically
optimized reactors are significant improved compared to the simple
2D reactors. To investigate the nature behind this improvement, we
show in Fig.~\ref{fig:PhiCompare} a log-log plot of the flow rate
$Q$ and the conversion $\conv$ normalized by the values
$\conv_{\mathrm{unif}}$ and $Q_{\mathrm{unif}}$ of the uniform
reactors at the same parameters. Because
Eq.~(\ref{eq:1DMeanRate2}) gives the following scaling of the
objective function $\Phi \sim Q\,\conv$, the rate of improvement
with respect to the uniform reactors can be read off directly, as
the contours of the improvement-factors of $\Phi$ become straight
lines, as showed by the dashed lines labelled by the corresponding
factors in Fig.~\ref{fig:PhiCompare}. It is seen that topology
optimization can increase the reaction rate of the optimal
reactors by nearly a factor 20, and furthermore it does so by
increasing the flow rate at the expense of lower conversions. The
important insight thus gained is that the distribution of the
advected reactant by the microfluidic channel network over a large
area at minimal pressure-loss plays a significant role when
optimizing microreactors.

To conclude, we have analyzed a single first-order catalytic
reaction in a laminar microfluidic reactor with optimal
distribution of porous catalyst. The flow is pressure-driven and
the flow through the porous medium is modelled using a simple
Darcy damping force.
%
Our goal has solely been to optimize the average reaction rate,
with no constrains on the conversion or the catalytic properties.
A characterization of the optimal configuration has been derived
theoretically and validated numerically. It shows an general
scaling behavior, depending only on the reaction properties of the
catalyst.
The analysis is based on a very simple reaction since this
emphasizes the points that the optimization of even simple
reactions result in to non-trivial scaling properties and complex
optimal designs.
Using topology optimization to design optimal reactors give rise
to reaction rate improvements of close to a factor 20, compared to
an corresponding optimal uniform reactor, and the improvement
originates mainly due to an improved transport and distribution of
the reactant. Furthermore, for the topology optimized reactors, we
have found a systematic decrease of pore-sizes and the emergence
of finer structures as the scaling parameter is decreased. Our
work points out a new, general, and potentially very powerful
method of improving microfluidic reactors.

F.~O.~was supported
by The Danish Technical Research Council No.~26-03-0037 and
No.~26-03-0073.

\end{document}